# ANGULAR DISTRIBUTION OF EXTENSIVE AIR SHOWERS BY TEL ARRAY UNDER GELATICA EXPERIMENT

Manana Svanidze[1], Yuri Verbetsky[1], Ia Iashvili[2], Abesalom Iashvili[1], Alexi Gongadze[1], Levan Kakabadze[1], George Kapanadze[1], Edisher Tskhadadze[1], George Chonishvili[3]

1. E Andronikashvili Institute of Physics under Tbilisi State University, Tbilisi, Georgia
2. The State University of New York at Buffalo, USA
3. J. Gogebashvili Telavi State University, Telavi, Georgia

***Abstract:*** *Extensive Air Showers' (EAS) arrival direction distribution is studied by means of a 4-detector installation in Telavi (TEL array), which is a node of GELATICA net in Georgia. The description of EAS arrival zenith angle distribution within the spheric layer model of the atmosphere and exponential absorption of showers with the air path is used. It is shown that the variation of zenith angles' upper cutoff boundary allows a stable estimation of showers' absorption path.*

***Keywords:*** *Extensive Air Showers, Angular Distribution, Resolution Function.*

## 1. Introduction

Arrival zenith angle distribution for the Extensive Air Showers (EAS) with a wide range of number of charged particles is studied using the experimental data obtained using small 4-detector array arranged under the concrete roof of main building of J.Gogebashvili Telavi State University (EAS goniometer TEL). The station is a part of the GELATICA net in Georgia (GEorgian Large-area Angle and Time Coincidence Array [1-3]). This long-term experiment is devoted to the study of possible correlations in the arrival time and direction of separate EAS events over large distances [4] and to the investigation of the Primary Cosmic Ray energy spectrum at very high energies.

The process of EAS development in the atmosphere with accompanied absorption manifests itself through the arrival direction distribution. That is why an interest to such investigations is long-standing [5-11]. The distribution of zenith angle $\theta$ of the shower arrival direction is usually studied under the assumption of azimuth isotropy for both the Cosmic Ray phenomenon and the measuring equipment.

It has been shown previously [10], that the distribution of zenith angle weakly depends on the energy of Primary Cosmic Ray particles. This feature makes it possible to investigate the subject, even by our small installations incapable of EAS energy direct measurement.

## 2. Description of the Installation

The TEL installation includes 4 scintillator detectors controlled by the data acquisition (DAQ) card [12], operating under PC control with a LabView interface for Windows (EAS goniometer). Detectors are arranged (Fig. 1) under the concrete roof and surrounded by concrete walls. The building is oblong nearly in the Southeast–Northwest direction.

Each detector consists of a 5 *cm* thick scintillators slab of (50×50) $cm^2$ area, supplied with a photo-multiplier tube (PMT). The PMT pulses, initiated by the passage of EAS charged particles through the scintillator material, are read by DAQ

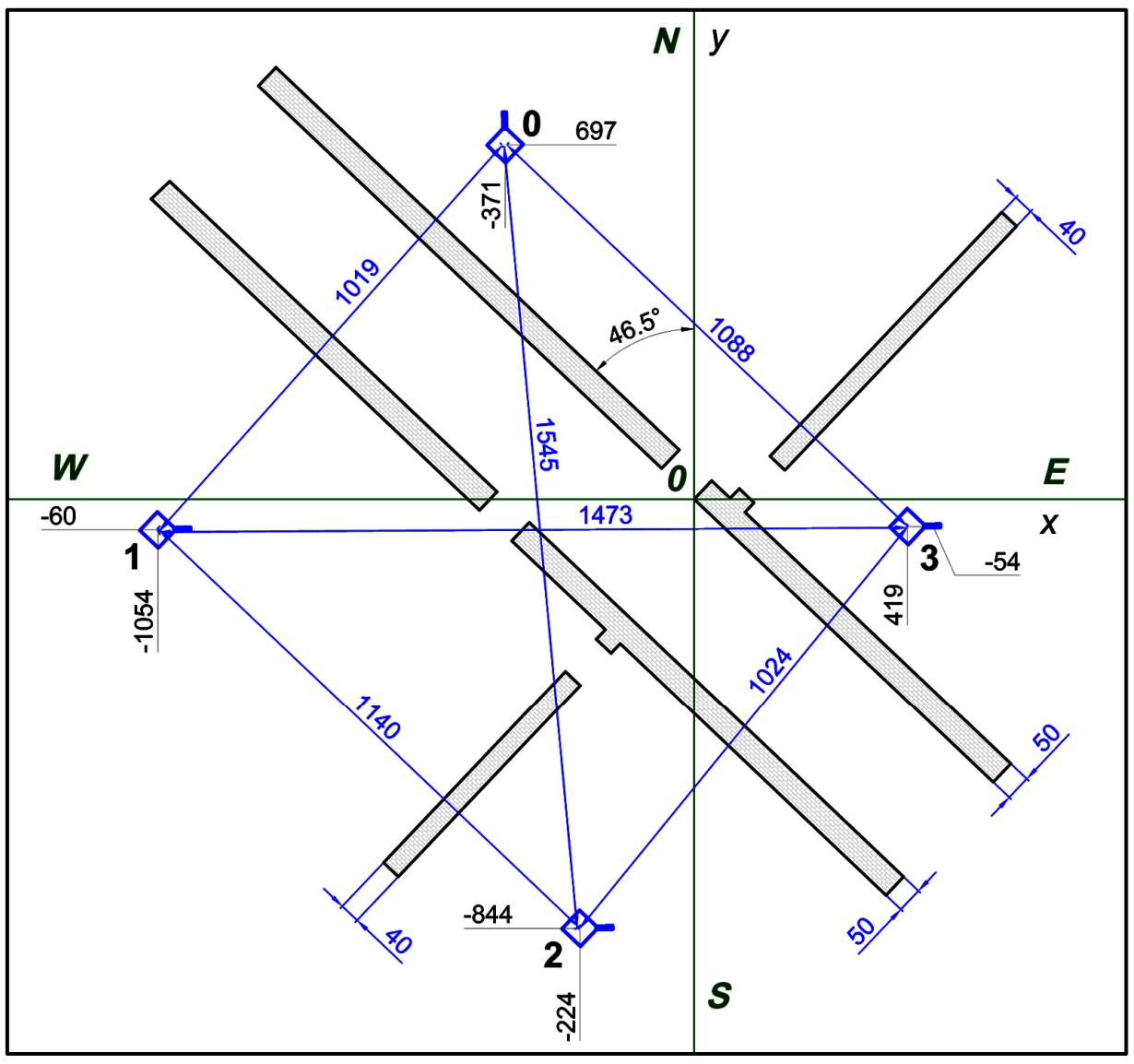

**Figure 1** TEL array layout. The gray rectangles display the horizontal profiles of the brick walls in the roof space used; the numerals **0, 1, 2, 3** are the detectors' labels; dimensions and coordinates are shown in centimeters relative to the East-North reference frame

card. The equipment measures the pulse delay relative to the 4-fold pulse coincidence with $\tau = 1.25$ *ns* time slicing step. The data are stored on PC in the form of integer values $k_0$, $k_1$, $k_2$ and $k_3$, corresponding to the numbers of delay slices for every detector shown on Fig. 1. This information allows a posterior estimation [13] of local direction of EAS front arrival.

## 3. Directions Estimation by Planar EAS Goniometer

The TEL installation is a planar EAS goniometer, permitting the linear estimation [13] of the planar (horizontal) components of the unit direction vector $\mathbf{n}$ of the EAS front's local tangent plane. It is assumed that the front of shower is moving with light velocity $c$.

Let us define a set of 2-vectors of the detectors' positions in horizontal plane in accordance with Fig. 1 as a matrix:

$$\mathrm{r} = \begin{pmatrix} x_0 & x_1 & x_2 & x_3 \\ y_0 & y_1 & y_2 & y_3 \end{pmatrix}. \tag{3.1}$$

The set of measured delay slices' numbers for every detector, corresponding to any EAS event observed is defined as:

$$\mathrm{k} = \begin{pmatrix} k_0 & k_1 & k_2 & k_3 \end{pmatrix}^{\mathrm{T}} \tag{3.2}$$

It is handy to shift all coordinates and slices' numbers by their averages; this shift does not change the EAS arrival direction estimation:

$$\bar{x} = \frac{1}{4}\sum_{n=0}^{3} x_n,\ \bar{y} = \frac{1}{4}\sum_{n=0}^{3} y_n,\ \bar{k} = \frac{1}{4}\sum_{n=0}^{3} k_n;$$

$$\mathrm{R} = \begin{pmatrix} x_0 - \bar{x} & x_1 - \bar{x} & x_2 - \bar{x} & x_3 - \bar{x} \\ y_0 - \bar{y} & y_1 - \bar{y} & y_2 - \bar{y} & y_3 - \bar{y} \end{pmatrix}, \tag{3.3}$$

$$\mathrm{K} = c\tau \cdot \begin{pmatrix} k_0 - \bar{k} & k_1 - \bar{k} & k_2 - \bar{k} & k_3 - \bar{k} \end{pmatrix}^{\mathrm{T}}. \tag{3.4}$$

Here the length $c\tau$ of the shower's front displacement during the DAQ's single slice is considered.

It is shown previously [13] that arrival direction ort horizontal components can be estimated by the linear least square method [14]. The correspondent normal equations are linear in our case:

$$\mathbf{A} \cdot \mathbf{n} = \mathbf{p} \tag{3.5}$$

Here the equation system matrix depends on the detectors' location only:

$$\mathbf{A} = \mathrm{R} \cdot \mathrm{R}^{\mathrm{T}} \tag{3.6}$$

while the right-hand term depends both on the detectors' location and on the measured delay slices' numbers:

$$\mathbf{p} = \mathrm{R} \cdot \mathrm{K} \tag{3.7}$$

Hence it follows that the estimation of the direction vector $\mathbf{n}$ is linear with respect to measured delay slices' numbers:

$$\mathbf{n} = \mathbf{A}^{-1} \cdot \mathbf{p} = \left(\mathbf{A}^{-1} \cdot \mathrm{R}\right) \cdot \mathrm{K} \tag{3.8}$$

Under the condition of identical properties of all detectors used, the dispersion matrix of the ort coordinates estimations (3.8) gets the form:

$$\mathbf{D}_0 = \sigma_0^2 \cdot \mathbf{A}^{-1} \tag{3.9}$$

The overall estimation of the front plane position dispersion is used:

$$\sigma_0^2 = \frac{1}{N_{\mathrm{det}} - N_{\mathrm{ort}} - N_{\mathrm{con}}}\left(\mathrm{K} - \mathrm{R}^{\mathrm{T}} \cdot \mathbf{n}\right)^2 \tag{3.10}$$

Here special numbers of degrees of freedom are used:

- number of measured delay slices (i.e. the number of detectors) $N_{\mathrm{det}} = 4$;
- number of the ort coordinates estimated $N_{\mathrm{ort}} = 2$;
- number of linear constraints due to delay slices shift $N_{\mathrm{con}} = 1$

The linear calculation technique stated here makes it possible to evaluate every EAS event data to get the values of EAS arrival ort horizontal components estimation both with the dispersion matrix of these components. Certainly, there exist some else sources of fluctuation of ort components' estimation, i.e. variation of the passage position of the triggering particle in every detector, uncertainty of the detectors' locations measurements, etc. The correspondent additional dispersions prove to be of considerably less significance then the received (3.10) main one. Though, these correcting dispersion matrixes are applied to the processing of TEL installation data.

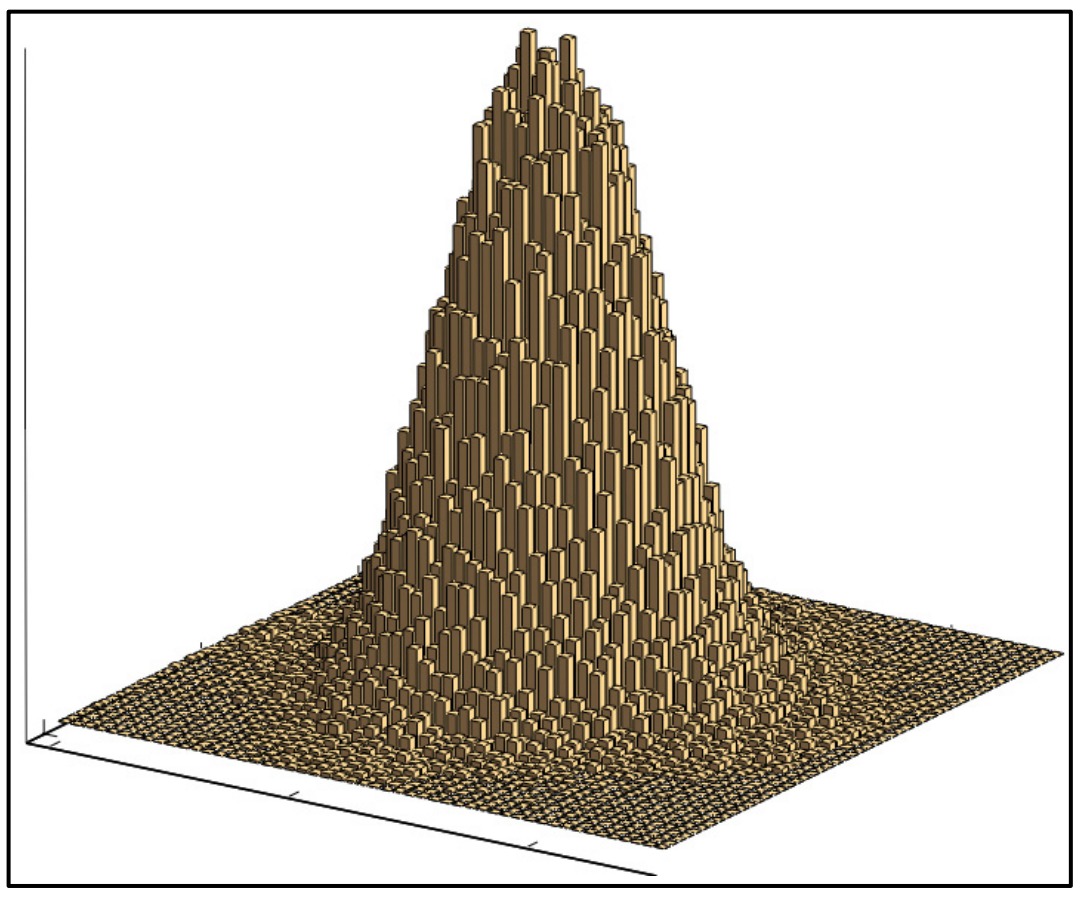

**Figure 2** Histogram of ($n_E$, $n_N$) components of EAS arrival direction orts by TEL installation data in the East-North reference frame

The distribution histogram of arrival direction ort components ($n_E$, $n_N$) of 151 671 EAS events (observed by TEL installation in the course of 5225 *hr* approximately) is shown at the Fig. 2. It represents the data analyzed hereinafter.

**4. TEL array location and respective model of spheric layer atmosphere used**

The consistent investigation of EAS arrival directions' distribution as a final goal of this study needs some reliable description of EAS absorption by the air surrounding the TEL installation, i.e. some reasonable model of the atmosphere.

The TEL installation is located at the altitude of $h_{\mathrm{TEL}} = (840 \pm 6)\,m$ by GPS estimation. The respective estimation of the installation vertical depth in the atmosphere

$$d_{\mathrm{TEL}}^{\mathrm{V}} = (936.8 \pm 0.7)\,g/cm^2 \tag{4.1}$$

is derived in accordance with International Civil Aviation Organization (ICAO) standard atmosphere parameters [15]. This value is used as one of the parameters describing the depth of the atmosphere along the view axis directed with angle $\theta$ relative to the zenith direction. The "flat" atmosphere model is used commonly [5, 8-11]. It exploits the conception of atmosphere as a flat layer of air with limited depth. This model suggests the dependence of the form

$$d_{\mathrm{p}}^{(\mathrm{flat})}(\theta) = d_{\mathrm{p}}^{\mathrm{V}} / \cos(\theta)\,. \tag{4.2}$$

Here $d_{\mathrm{p}}^{\mathrm{V}}$ is the vertical depth of atmosphere in the investigation point location.

Certainly $d_{\mathrm{p}}^{\mathrm{V}} = d_{\mathrm{TEL}}^{\mathrm{V}}$ in our case.

It is obvious that the air depth calculated according to (4.2) grows unrestrictedly in the horizon vicinity. Only directions closer then 60° to zenith direction are usually allowed for cosmic ray absorption studies in this model.

Our previous investigation of EAS arrival directions' distribution based on the data of TBS installation in Tbilisi [11] has revealed the significance of the EAS flux consideration until the 70° value of Zenith angle. That is why we use somewhat more sophisticated model of "spheric layer" atmosphere in this study. It is a common geometric calculation to get the dependence

$$d_{\mathrm{p}}^{(\mathrm{sph})}(\theta) = d_{\mathrm{p}}^{\mathrm{V}} \frac{2 + C_{\mathrm{p}}}{\cos(\theta) + \sqrt{\left(1 + C_{\mathrm{p}}\right)^2 - \sin^2(\theta)}} \tag{4.3}$$

for the air depth along the view axis directed with zenith angle $\theta$ – for the atmosphere imagined as a spheric layer with limited vertical depth. Here the location-specific parameter $C_{\mathrm{p}}$ describes the exploration point location in relation with the Earth globe dimensions. The air depth (4.3) in the spheric layer model in the evident limit $C_{\mathrm{p}} \to 0$ (i.e. as if the radius of the glob tends to infinity) tends to the form (4.2) of the flat atmosphere model.

The exploration-point specific parameter $C_{\rm p}$ is unknown. It may be determined by comparison of the spheric layer model of the atmosphere prediction (4.3) with the prediction of ICAO standard model of the atmosphere (taken as the reference model). The air depth along the view axis in this model is estimated as integral of air density along the view ray. The parameter $C_{\rm st}$ value may be estimated by minimization of average square relative difference between the predictions of the two models. This method in our case results in value

$$C_{\rm TEL} = 2.04095 \times 10^{-3} \tag{4.4}$$

The required limit of spheric layer model applicability $\theta_{\max}$ is estimated by study of the relative deviation between the spheric layer and ICAO standard atmosphere models:

$$Q_{\rm TEL}(\theta) = \frac{d_{\rm TEL}^{\rm (sph)}(\theta)}{d_{\rm TEL}^{\rm (ICAO)}(\theta)} - 1 .$$

We require the spheric layer model accuracy restricting this value by limits $\left|Q_{\rm TEL}(\theta)\right| < 0.5\%$. One may ascertain from the Fig. 3 that the spheric layer atmosphere model is satisfactory one for zenith angles $0 < \theta < \theta_{\max}$, $\theta_{\max} = 87.2°$ in the case of TEL installation. The flat atmosphere model becomes unacceptable for much lower zenith angles.

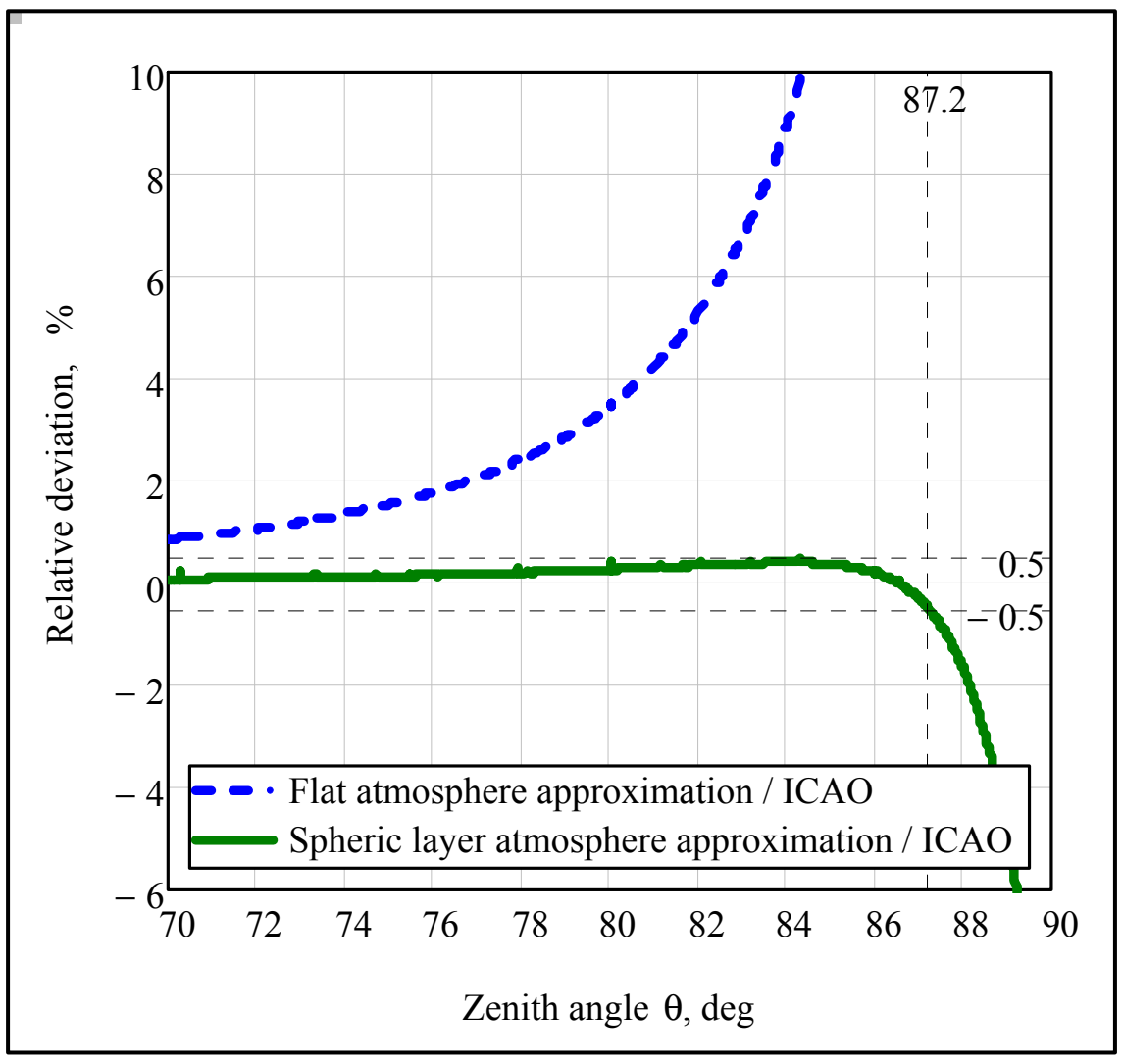

**Figure 3** Relative deviations of two layer models of the atmosphere with respect to ICAO standard one for TEL installation location

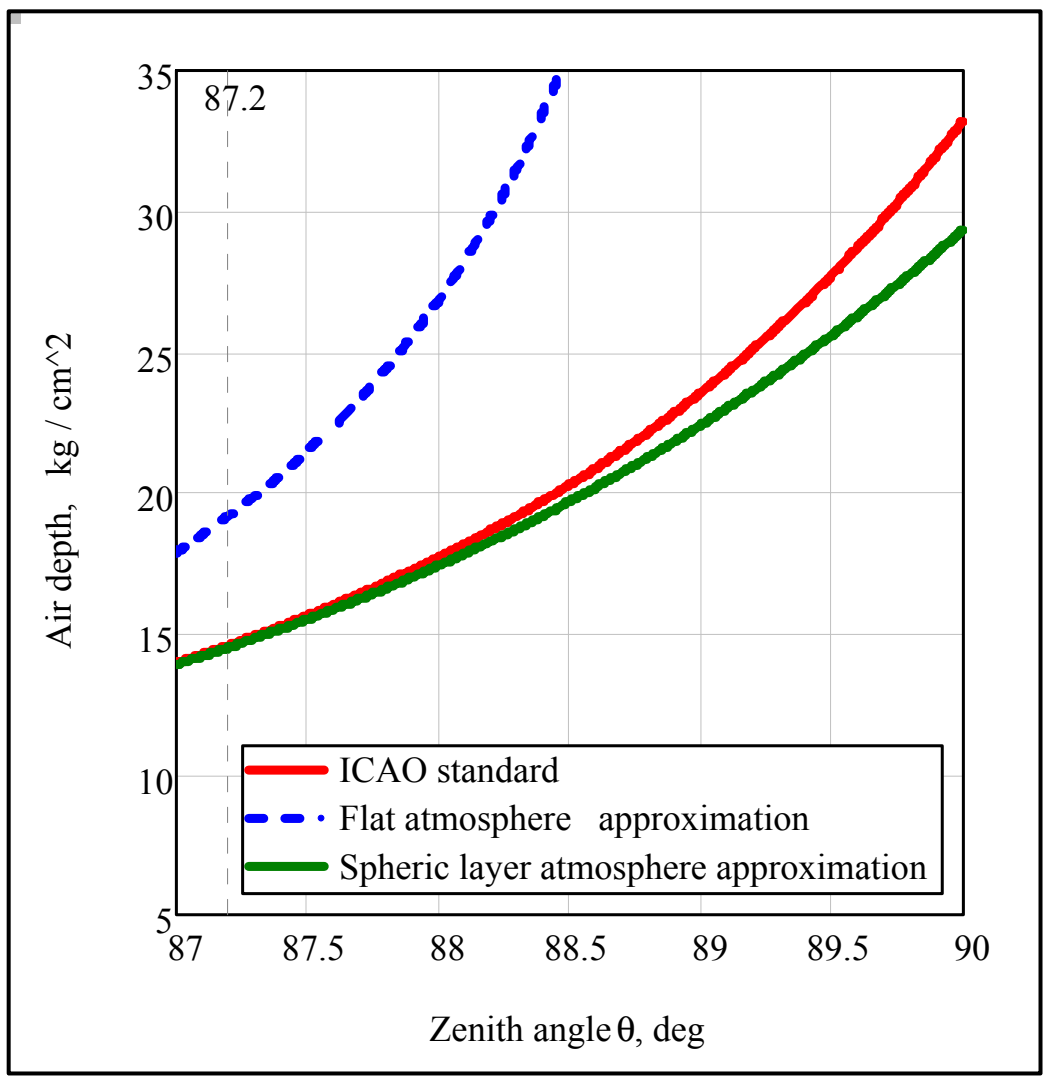

**Figure 4** Comparison of the air depth predictions by the three models of the atmosphere in the horizon vicinity for TEL installation location

The air depth predictions in the horizon vicinity by the three models examined are shown in Fig. 4. In contrast to the flat atmosphere model, the spheric layer one underestimates slightly the total air depth in horizontal direction

So, the spheric layer atmosphere model proposed satisfies applicability requirement for our study for zenith angles $\theta \le 87°$. The air depth dependence on the zenith angle (4.3) is analytically simple in contrast to the awkward numerical integration of air density computed by ICAO model and can be used in the subsequent application for EAS arrival direction distribution.

## 5. Considered Form of Distribution

As is mentioned above, the planar goniometers are capable of straight estimation of two components of EAS arrival direction unit vector only, i.e. ($n_{\rm E}$, $n_{\rm N}$), being parallel to the detectors' location plane [13]. That is why the immediate variable, independent of any additional assumption and measuring the event direction separation from the zenith direction is the estimated length of the unit direction vector projection onto the detectors' plane

$$\beta = \sqrt{n_{\rm E}^2 + n_{\rm N}^2} \tag{5.1}$$

This variable gives estimate of the usual zenith angle indirectly.

The correspondent geometric zenith separation variable

$$\alpha = \sin(\theta) \tag{5.2}$$

is restricted to the finite interval $0 \le \alpha \le 1$, while the estimated value $\beta$ of the events' zenith separation may exceed the geometric limit of unity due to estimation (5.1) errors.

### *5.1. Fundamental Distribution of Zenith Separation*

We shall assume that all EAS developed in the atmosphere are absorbed in compliance with the usual exponential low [10]. Thus the flux of EAS observed in the unit solid angle by the installation located at a point under the air depth $d_{\mathrm{p}}(\theta)$ is proportional to

$$\exp\left\{-d_{\mathrm{p}}(\theta)\big/\Lambda_{\mathrm{abs}}\right\} \tag{5.3}$$

Here $\Lambda_{\mathrm{abs}}$ is the EAS absorption path required. The dependence is assumed to be applicable in the framework of a spheric layer atmosphere model (4.3), i.e. in the interval $0 \le \theta \le 87°$ of zenith angles. Taking into consideration that the TEL goniometer employs the flat detectors located in the horizontal plane (i.e. adding a $\cos(\theta)$ factor to (5.3)), let us integrate the observed flux expression by the azimuth to get a zenith angle distribution in the form of

$$\sin(\theta)\cos(\theta)\exp\left\{-d_{\mathrm{p}}^{(\mathrm{sph})}(\theta)\big/\Lambda_{\mathrm{abs}}\right\} \tag{5.4}$$

The distribution of EAS arrival zenith angle (5.4) expressed in the terms (5.2) of zenith separation variable $\alpha$ proves to be the function:

$$f_{\alpha}(\alpha\,|\,q) = \frac{\alpha\cdot\exp\left(-q\dfrac{2+C_{\mathrm{p}}}{\sqrt{1-\alpha^2}+\sqrt{\left(1+C_{\mathrm{p}}\right)^2-\alpha^2}}\right)}{\displaystyle\int_0^1 \xi\cdot\exp\left(-q\frac{2+C_{\mathrm{p}}}{\sqrt{1-\xi^2}+\sqrt{\left(1+C_{\mathrm{p}}\right)^2-\xi^2}}\right)d\xi} \tag{5.5}$$

Here value $q = d_{\mathrm{p}}^{\mathrm{V}}\big/\Lambda_{\mathrm{abs}}$ measures the EAS absorption range number in vertical direction. The fundamental distribution function (5.5) has to be corrected to take into account the distortions by the surrounding matter anisotropy and installation's resolution function before the comparison with TEL data in the following.

In contrast to the previous investigation of EAS arrival directions' distribution [11] by means of TBS installation data, the TEL installation data under consideration proves the anisotropy account to be negligible. That is why only the influence of installation's resolution function is considered further.

### *5.2. Resolution Function*

Detectors of TEL installation are located almost symmetrically at the vertices of a square. The estimations of components of EAS arrival direction vector are almost uncorrelated and equal-dispersion in this case. The ort components' estimations are obtained by means of linear transformation (3.8) of directly measured integer timing $k$ numbers of signals' from the detectors. Therefore it is possible to use the assumption that the joint distribution of estimates of ($n_{\mathrm{E}}$, $n_{\mathrm{N}}$) components can be approximated by the Normal distribution, with the approximate dispersion matrix $\widetilde{\mathbf{D}}$ proportional to identity:

$$G(\mathbf{n}\,|\,\mathbf{n}_0,\sigma) \sim \exp\left\{-\frac{1}{2}(\mathbf{n}-\mathbf{n}_0)^{\mathrm{T}}\cdot\widetilde{\mathbf{D}}^{-1}\cdot(\mathbf{n}-\mathbf{n}_0)\right\};\qquad \mathbf{n} = \begin{pmatrix} n_{\mathrm{E}} \\ n_{\mathrm{N}} \end{pmatrix};\qquad \widetilde{\mathbf{D}} = \begin{pmatrix} \sigma^2 & 0 \\ 0 & \sigma^2 \end{pmatrix};$$

Here $\mathbf{n}_0$ vector represents the position of true direction (unknown). Only $\mathbf{n}$ vector may be measured, with $\widetilde{\mathbf{D}}$ uncertainty, of course. The averaged dispersion $\sigma^2$ is defined further.

Let us integrate this distribution by azimuth to obtain the radial distribution of measured zenith separation $\beta$ needed.

While $\beta$ is a measured separation (5.1) from EAS arrival direction to the zenith direction and $\varphi = \arctan(n_{\mathrm{N}}/n_{\mathrm{E}})$ is the correspondent azimuth angle, we get the expression in polar coordinate system:

$$\frac{1}{2}(\mathbf{n}-\mathbf{n}_0)^{\mathrm{T}} \cdot \widetilde{\mathbf{D}}^{-1} \cdot (\mathbf{n}-\mathbf{n}_0) = \frac{1}{2\sigma^2}\left[\alpha^2 + \beta^2 - 2\alpha\beta\cos(\varphi-\varphi_0)\right].$$

The azimuth integral of the correspondent exponent is:

$$\int_0^{2\pi} \mathrm{e}^{-\frac{1}{2\sigma^2}\left[\alpha^2+\beta^2-2\alpha\beta\cos(\varphi-\varphi_0)\right]} d\varphi = 2\pi \cdot \mathrm{e}^{-\frac{\alpha^2+\beta^2}{2\sigma^2}} \cdot \mathrm{I}_0\left(\frac{\alpha\beta}{\sigma^2}\right)$$

It is handy to define a scaled Bessel function $\mathrm{I}_0^*(x) = \mathrm{e}^{-x}\,\mathrm{I}_0(x)$ to get useful form of further expressions.

Consequently the resolution function, i.e. conditional distribution of unbounded $\beta \geq 0$ measured variable estimation under the assumption, that $\alpha$ is the true value of this separation, can be defined as:

$$\mathrm{Res}(\beta\,|\,\alpha,\sigma^2) = \frac{\beta}{\sigma^2}\mathrm{e}^{-\frac{(\alpha-\beta)^2}{2\sigma^2}} \cdot \mathrm{I}_0^*\left(\frac{\alpha\beta}{\sigma^2}\right); \qquad \int_0^{\infty} \mathrm{Res}(\beta\,|\,\alpha,\sigma^2)\,d\beta = 1. \tag{5.6}$$

So, it is necessary to employ some previously obtained averaged estimation of dispersions $\sigma^2$ of measured orts' components. Let us define these averaged dispersions for every EAS event as:

$$\sigma^2 = \sqrt{\det(\mathbf{D})}; \quad \mathbf{D} = \begin{pmatrix} \sigma_{\mathrm{E}}^2 & \sigma_{\mathrm{EN}} \\ \sigma_{\mathrm{EN}} & \sigma_{\mathrm{N}}^2 \end{pmatrix}$$

Here matrix $\mathbf{D}$ is a complete dispersion matrix of ($n_{\mathrm{E}}$, $n_{\mathrm{N}}$) components estimation for separate EAS event. The observed data allows determination of the averaged dispersion $\sigma^2$ dependence on the event's separation from zenith. This dependence is approximated by some regression polynomial $\sigma^2(\beta)$, as is shown in Fig. 5. It is used for resolution functions (5.6) construction. Figure 6 shows this function for some values of true zenith separation.

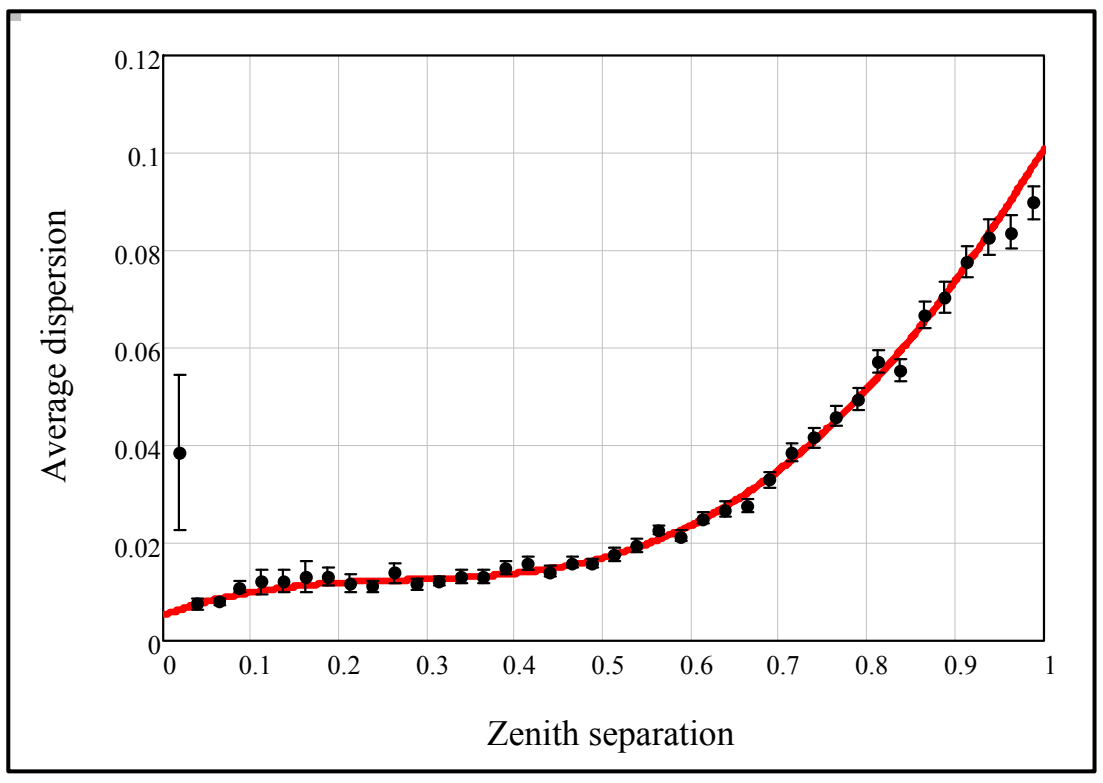

**Figure 5** Dependence of average dispersion $\sigma^2$ of ort components' estimation on measured zenith separation $\beta$

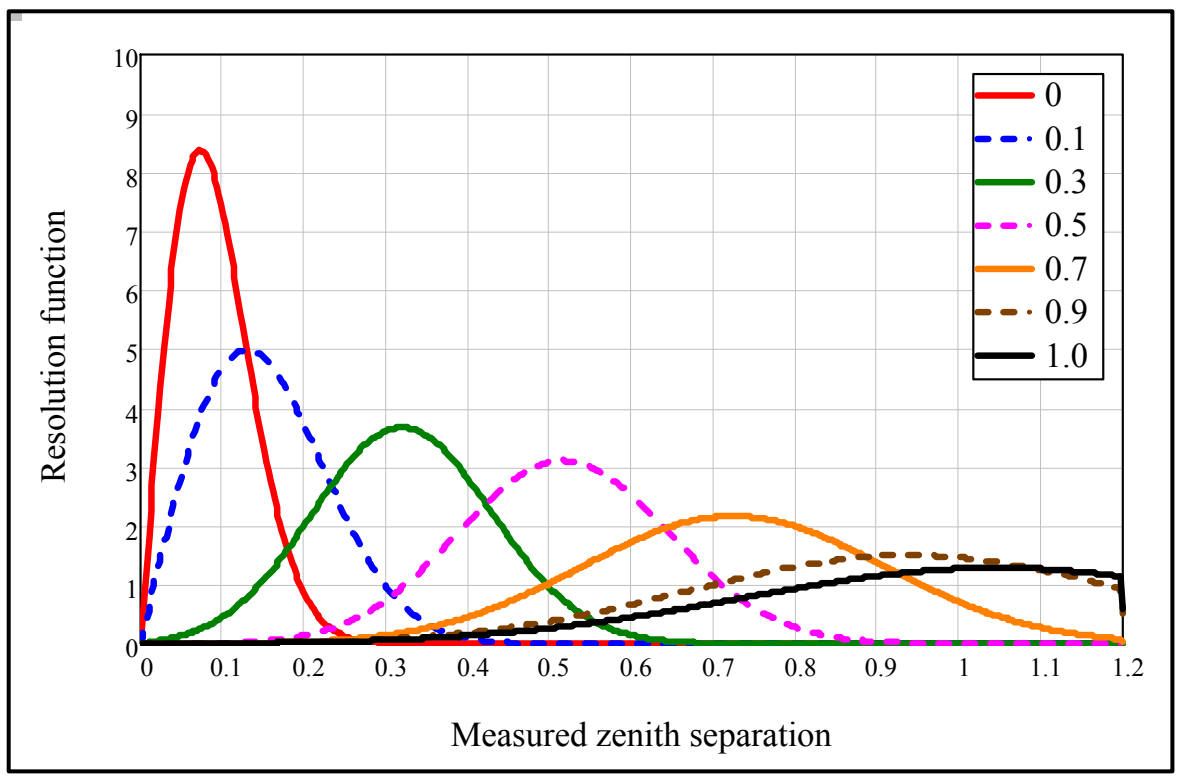

**Figure 6** Resolution functions $\mathrm{Res}(\beta\,|\,\alpha,\sigma^2(\alpha))$ for some specified values of true zenith separation $\alpha$

The definitions (5.5) and (5.6) now allow constructing the final probability density function of the event's observed separation $\beta$ from zenith direction:

$$f_\beta(\beta\,|\,q) = \int_0^1 \mathrm{Res}(\beta\,|\,\alpha,\sigma^2(\alpha)) \cdot f_\alpha(\alpha\,|\,q)\,d\alpha \tag{5.7}$$

This distorted density function is the one to be compared with the TEL observed data to estimate the $q$ value, i.e. the EAS absorption range number in vertical direction.

## 6. Estimation of the EAS absorption range

The method of maximum likelihood is used for this purpose, applied to the part of the data in the angular interval of $0 \le \theta \le 87.2°$ or $0 \le \beta \le B_{max}$, $B_{max} = 0.9988$, i.e. in the range of applicability of the spheric layer atmosphere model (4.3) for the air depth along the view axis.

Let us investigate dependence of the EAS absorption vertical range number $q$ value on the position of the upper truncation limit $\beta_{tr}$ of the data subset used. The correspondent truncated versions of distorted density function (5.7) are used in likelihood construction for comparison with the TEL truncated data subsets. The sequence of $q$ estimations has been obtained by repeatedly applying the maximum likelihood method to the sequentially expanding data subsets. The resulting dependence is shown in Fig. 7. (All points in this sequence of estimations are mutually dependent!)

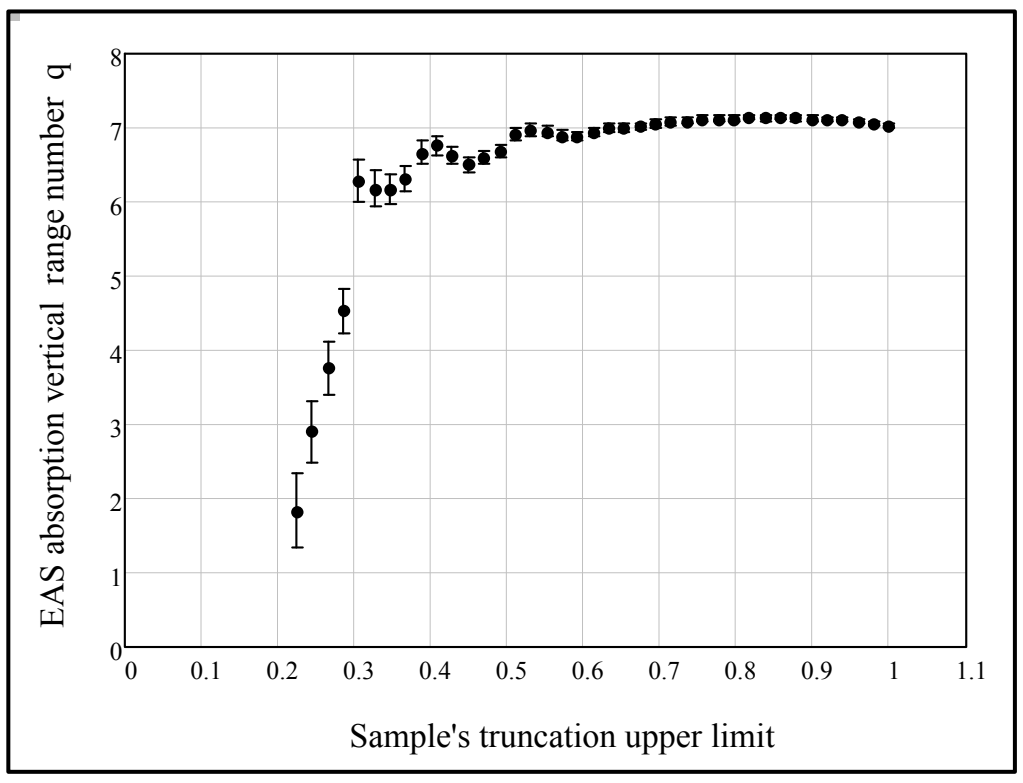

**Figure 7** Dependence the EAS absorption vertical range number $q$ on the sample's truncation limit $\beta_{tr}$

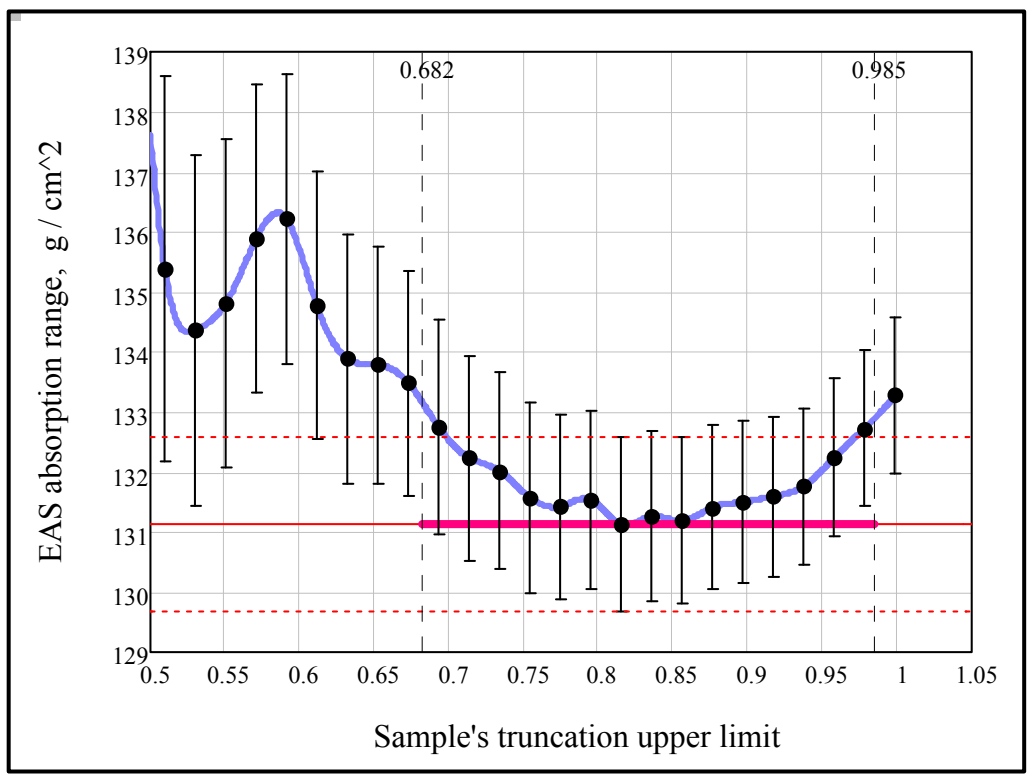

**Figure 8** Dependence of EAS absorption range estimation on the sample's truncation limit $\beta_{tr}$

This result allows estimation of the required EAS absorption range $\Lambda_{abs} = d^{V}_{TEL}/q$ dependence on the positions of the upper truncation limits $\beta_{tr}$ of data subsets used. The resulting dependence is shown in Fig. 8 for the interval of certain stabilization. As can be seen from the last figure, these estimations become stable within one standard deviation width for the truncation limits within the interval of $0.682 \le \beta_{tr} \le 0.985$. That is why we adopt the final estimation of EAS absorption path:

$$\Lambda^{(TEL)}_{abs} = (131.1 \pm 1.4)\, g/cm^2 ,$$

corresponding to the maximal width of the interval of stability. It is valid within the interval $0 \le \alpha \le 0.985$ of event's separation from zenith, therefore. The correspondent upper limiting zenith angle is approximately 80°.

The installation's resolution broadens the distribution of existing data as compared with correspondent fundamental physical distribution.

The key influence of this distorting impact is explicitly expressed in the difference of the fundamental distribution (5.5) and fitted distorted one (5.7), shown in Fig. 9. Any incautious attempt of immediate fitting of fundamental distribution to the existing data results in unstable estimation of $\Lambda_{abs}$, not in agreement with existing previous world data.

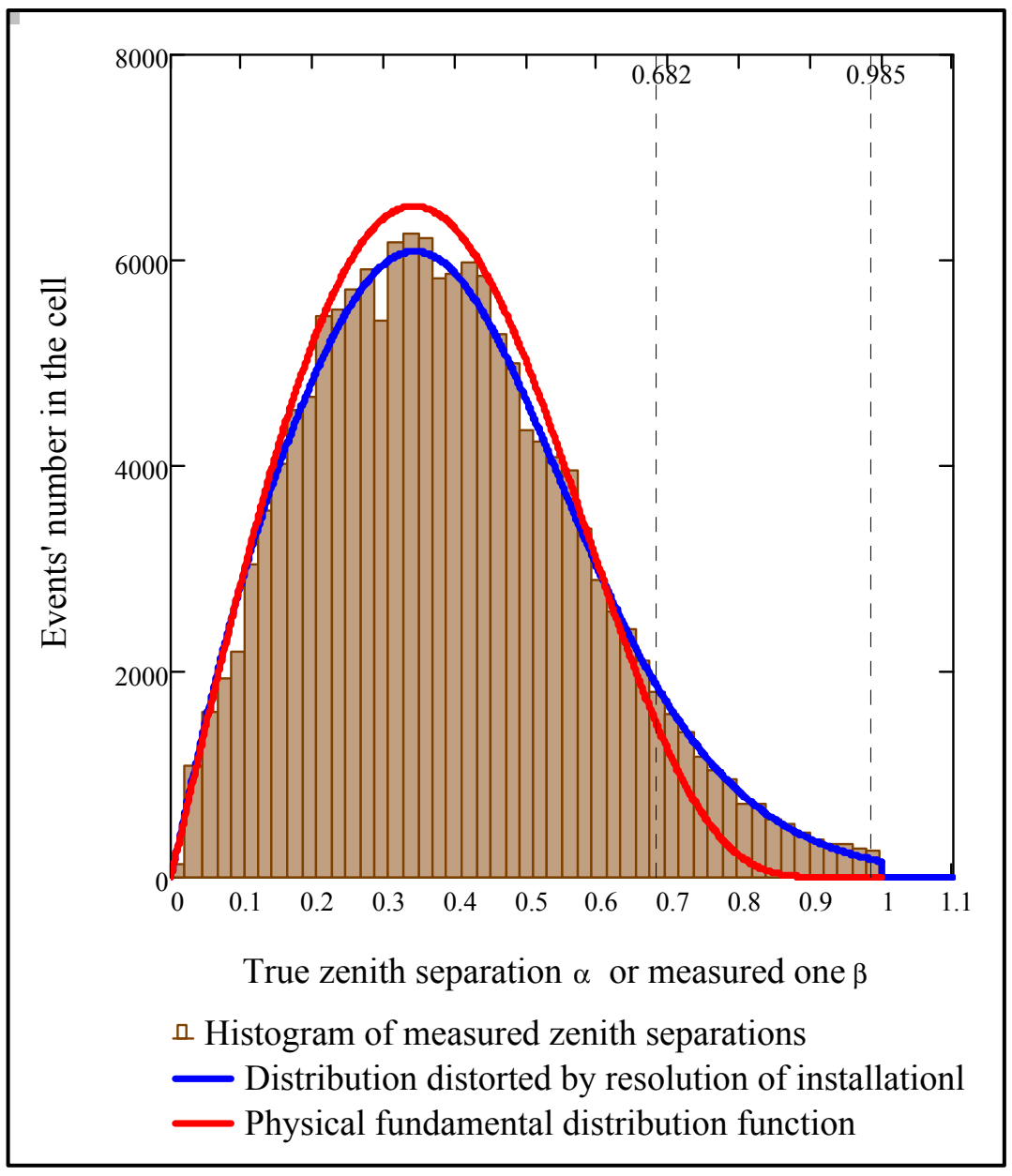

**Figure 9** Comparison of the observed data histogram and distributions obtained

## 7. Conclusions

It has been established that taking into account for the resolution function of TEL installation makes it possible to employ the fundamental distribution (5.5) of event's separation from zenith $\alpha = \sin(\theta)$. This model of regular EAS absorption, in accordance with spheric layer atmosphere model approximation (4.3), has proved to be valid for description of EAS absorption process within the interval $0 \le \alpha \le 0.985$ of event's separation from zenith, i.e. in the interval $0 \le \theta \le 80°$ of zenith angle. The estimated value of EAS absorption path is actually stable under variation of data truncation upper limits within $0.682 \le \beta_{\mathrm{tr}} \le 0.985$ and is fixed at the magnitude $\Lambda_{\mathrm{abs}}^{(\mathrm{TEL})} = (131.1 \pm 1.4)\, g/cm^2$. Any estimation of this parameter upon the more restricted sequence of intervals of $\beta$ variable is unstable. It is the immediate consequence of this study, that any attempt of absorption path estimation with use of some data truncation, not proved to be consistent with stability under variation of this truncation limit, is unreliable, indeed. Our $\Lambda_{\mathrm{abs}}^{(\mathrm{TEL})}$ estimation is in approximate agreement with the previous estimations by installations located at various altitudes:

$$\Lambda_{\mathrm{abs}} = (135 \pm 10)\, g/cm^2 \text{ [5]}, \quad \Lambda_{\mathrm{abs}} = (130 \pm 7)\, g/cm^2 \text{ [7]}, \quad \Lambda_{\mathrm{abs}} = (106 \pm 6)\, g/cm^2 \text{ [9]},$$

$$\Lambda_{\mathrm{abs}} = (115 \pm 0.5)\, g/cm^2 \text{ [10]}, \quad \Lambda_{\mathrm{abs}} = (115.4 \pm 2.6)\, g/cm^2 \text{ [11]}.$$

## Acknowledgements

The authors are grateful to other current and former members of our group for their technical support. We are especially thankful to our colleagues working now in foreign. This work was supported by the Georgian National Science Foundation subsidy for a grant of scientific researches #GNSF/ST06/4-075 (№ 356/07) and by Shota Rustaveli National Science Foundation, Project #DI/6/6-300/12.